\documentclass{pasj00}
\begin{document}
\SetRunningHead{Danezis et al.}{A new model for the structure of
the DACs and SACs regions in the Oe and Be stellar atmospheres}
\Received{2000/12/31}%{yyyy/mm/dd}
\Accepted{2001/01/01}%{yyyy/mm/dd}

\title{A new model for the structure of the DACs and SACs regions in the Oe
and Be stellar atmospheres}

%%% begin:list of authors
\author{Emmanouel \textsc{Danezis}%
%\thanks{Example: Present Address is xxxxxxxxxx}
}
\affil{University of Athens, Faculty of Physics Department of
        Astrophysics, Astronomy and Mechanics}
\affil{Panepistimioupoli, Zographou 157 84, Athens, Greece}
\email{edanezis@phys.uoa.gr}

\author{Dimitris \textsc{Nikolaidis}}
\affil{University of Athens, Faculty of Physics Department of
        Astrophysics, Astronomy and Mechanics}
\affil{Panepistimioupoli, Zographou 157 84, Athens, Greece}
\email{edanezis@phys.uoa.gr}

\author{Evaggelia \textsc{Lyratzi}}
\affil{University of Athens, Faculty of Physics Department of
        Astrophysics, Astronomy and Mechanics}
\affil{Panepistimioupoli, Zographou 157 84, Athens, Greece}
\email{elyratzi@phys.uoa.gr}

\author{Luka \v{C}. \textsc{Popovi\'{c}}}%
\affil{Astronomical Observatory of Belgrade, Volgina 7, 11160
        Belgrade, Serbia}
\affil{Isaac Newton Institute of Chile, Yugoslavia
        Branch}
\email{lpopovic@aob.bg.ac.yu}

\author{Milan S. \textsc{Dimitrijevi\'{c}}}%
\affil{Astronomical Observatory of Belgrade, Volgina 7, 11160
        Belgrade, Serbia}
\affil{Isaac Newton Institute of Chile, Yugoslavia
        Branch}
\email{mdimitrijevic@aob.bg.ac.yu}

\author{Antonis \textsc{Antoniou}}
\affil{University of Athens, Faculty of Physics Department of
        Astrophysics, Astronomy and Mechanics}
\affil{Panepistimioupoli, Zographou 157 84, Athens, Greece}
\email{ananton@phys.uoa.gr}

\and
\author{Efstratios {\sc Theodosiou}}
\affil{University of Athens, Faculty of Physics Department of
        Astrophysics, Astronomy and Mechanics}
\affil{Panepistimioupoli, Zographou 157 84, Athens, Greece}
\email{etheodos@phys.uoa.gr}
%%% end:list of authors

%%% Please use the following style in case that sorting by
%%% affilation is impossible.
%
% \author{%
%   Emmanouel \textsc{Danezis}\altaffilmark{1}
%   Dimitris \textsc{Nikolaidis}\altaffilmark{1}
%   Evaggelia \textsc{Lyratzi}\altaffilmark{1}
%   Luka \v{C}. \textsc{Popovi\'{c}}\altaffilmark{2,3}
%   Milan S. \textsc{Dimitrijevi\'{c}}\altaffilmark{2,3}
%    \textsc{}\altaffilmark{1}
%   Antonis \textsc{Antoniou}\altaffilmark{1}
%   and
%   Efstratios \textsc{Theodosiou}\altaffilmark{1}}
% \altaffiltext{1}{University of Athens, Faculty of Physics Department of
%        Astrophysics, Astronomy and Mechanics Panepistimioupoli,
%        Zographou 157 84, Athens, Greece}
% \email{edanezis@phys.uoa.gr}
% \altaffiltext{2}{Astronomical Observatory of Belgrade, Volgina 7, 11160
%        Belgrade, Serbia}
% \altaffiltext{3}{Isaac Newton Institute of Chile, Yugoslavia
%        Branch}

%% `\KeyWords{}' always has to be placed before `\maketitle'.
 %Do NOT move this preamble from here!

\maketitle

\begin{abstract}
In this paper we present a new mathematical model for the density
regions where a specific spectral line and its SACs/DACs are
created in the Oe and Be stellar atmospheres. In the calculations
of final spectral line function we consider that the main reasons
of the line broadening are the rotation of the density regions
creating the spectral line and its DACs/SACs, as well as the
random motions of the ions. This line function is able to
reproduce the spectral feature and it enables us to calculate some
important physical parameters, such as the rotational, the radial
and the random velocities, the Full Width at Half Maximum, the
Gaussian deviation, the optical depth, the column density and the
absorbed or emitted energy. Additionally, we can calculate the
percentage of the contribution of the rotational velocity and the
ions' random motions of the DACs/SACs regions to the line
broadening. Finally, we present two tests and three short
applications of the proposed model.
\end{abstract}

\section{Introduction}

When we study the UV lines of hot emission stars we have to deal
with two problems: (i) The presence of a very complex structure of
many spectral lines in the UV region, such as the resonance lines
of Si IV, C IV, N V, Mg II and the N IV spectral line. (ii) The
presence of Discrete Absorption Components (DACs, \citet{bat86}).

DACs are discrete but not unknown absorption spectral lines. They
are spectral lines of the same ion and the same wavelength as a
main spectral line, shifted at different $\Delta\lambda$, as they
are created in different density regions which rotate and move
radially with different velocities \citep{dan83, dan87, dan03,
lyr07}. DACs are lines, easily observed, in the case that the
regions that give rise to such lines, rotate with low velocities
and move radially with high velocities.

Many suggestions have been made in order to explain the DACs
phenomenon. Most of researchers have suggested mechanisms that
allow the existence of structures which cover all or a significant
part of the stellar disk, such as shells, blobs or puffs
\citep{und75, hen84, und84, bat86, gra87, lam88, wal92, wal94,
cra96, riv97, kap96, kap97, kap99, mar00}, interaction of fast and
slow wind components, Corotation Interaction Regions (CIRs),
structures due to magnetic fields or spiral streams as a result of
the stellar rotation \citep{und84, mul84a, mul84b, mul86, pri88,
cra96, ful97, kap96, kap97, kap99, cra00}. Though we do not know
yet the mechanism responsible for the formation of such
structures, it is positive that DACs result from independent high
density regions in the stellar environment.

An important question is whether there is a connection between the
observed complex structure of the above spectral lines and the
presence of DACs. A possible answer is that if the regions that
create the DACs rotate with large velocities and move radially
with small velocities, the produced lines have large widths and
small shifts. As a result, they are blended among themselves as
well as with the main spectral line and thus they are not
discrete. In such a case the name Discrete Absorption Components
is inappropriate and we use only the name Satellite Absorption
Components (SACs) \citep{lyr04, dan05, dan06, nik06a, nik06b}. A
very important question is whether it is possible that the
existence of SACs may be responsible for the complex structure of
the observed spectral feature.

As it is clear, for a future study of the mechanisms responsible
for the creation of the density regions which produce the complex
profiles of the above spectral lines, as well as for the study of
their structure and evolution, we need to calculate the values of
some physical parameters of these regions.

In order to calculate these parameters, it is necessary to
construct a line function, based on the idea of DACs and SACs
phenomena able to reproduce, in the best way, the observed
spectral feature.

\citet{dan03} presented such a line function in the case that the
reason of the line broadening is only the rotation of the regions
which create the observed spectral lines and their components and
when these regions present spherical symmetry around their own
center or the center of the star (see also \citet{lyr07}). With
this line function we calculate some important physical
parameters, such as the rotational and the radial velocities, the
optical depth, the column density \citep{dan05} and the
absorbed/emitted energy. But, in some cases the chaotic motion of
emitters/absorbers inside the dense layers can be significant (see
e.g. \citet{dan06}) and the estimated rotation might be
overestimated. Therefore, the modification of the previously given
model \citep{dan03} is needed, in sense that the possible
contribution of random velocities to the line broadening can be
taken into account.

Here we present a new model which includes also the random motions
of the ions, since in the case of hot emission stars one can
expect a significant contribution of random motion of the
emitters/absorbers to the line profile. Based on this idea,
besides the above mentioned physical parameters of the regions
that create the complex spectral lines, we can calculate the mean
random velocity of the ions, but also the percentage of the
contribution of the rotational velocity and emitter/absorber
random motions of the DACs/SACs regions to the line broadening, as
well as the Gaussian deviation for each DAC/SAC profile.

The paper is organized as follows: In \S 2 we give the description
of the model, in \S 3 we discuss the model, in \S 4 we give two
tests of the model, in \S 5 we present  applications of the model
and finally, in \S 6 we outline our conclusions.

\section{Description of the model - The Line Function}

As was already mentioned above, \citet{dan03} presented a line
function which is able to reproduce accurately the observed
spectral lines and their SACs/DACs in the same time. The proposed
line function is:

\begin{equation}
\label{1} F_{\lambda final} = {\left[ {F_{0} \left( {\lambda}
\right){\prod\limits_{i} {e^{ - L_{i} \xi _{i}}  +
{\sum\limits_{j} {S_{\lambda ej} \left( {1 - e^{ - L_{ej} \xi
_{ej}}}  \right)}}} }} \right]} e^{ - L_{g} \xi _{g}}
\end{equation}

\noindent where:

\noindent $F_{0} \left( {\lambda}  \right)$: the initial radiation
flow,

\noindent $L_{i}$, $L_{ej}$, $L_{g}$: are the distribution
functions of the absorption coefficients $k_{\lambda i}$,
$k_{\lambda ej}$, $k_{\lambda g}$ respectively. Each $L$ depends
on the values of the apparent rotational velocity as well as of
the radial velocity of the density shell, which forms the spectral
line ($V_{rot}$, $V_{rad}$)

\noindent $\xi_{i}$: is the optical depth in the center of the
line,

\noindent $S_{\lambda ej}$: the source function, which, at the
moment when the spectrum is taken, is constant.

In Eq.~\ref{1}, the functions $e^{ - L_{i} \xi _{i}} $,
$S_{\lambda ej} \left( {1 - e^{ - L_{ej} \xi _{ej}}}  \right)$,
$e^{ - L_{g} \xi _{g}} $ are the distribution functions of each
satellite component and we can replace them with a known
distribution function (Gauss, Lorentz, Voigt or disk model). An
important fact is that in the calculation of $F(\lambda)$ we can
include different geometries (in the calculation of $L$) of the
absorbing or emitting independent density layers of matter.

The decision on the geometry is essential for the calculation of
the distribution function that we use for each component, i.e. for
different geometries we have  different line shapes, representing
the considered SACs.

Eq.~\ref{1} gives the function of the complex profile of a
spectral line, which presents SACs or DACs. This means that
Eq.~\ref{1} is able to reproduce not only the main spectral line,
but its SACs as well.

The main hypotheses when we constructed the Rotation distribution
function were that the line width $\Delta\lambda $ is only due to
the rotation of the regions which create the observed spectral
lines and their components and that these regions present
spherical symmetry around their own center or the center of the
star (see also \citet{lyr07}). Consequently, the random velocities
of the emitters/absorbers in the density region are assumed to be
negligible, i.e. they do not significantly contribute to the line
profile. Here we consider that the random velocities may have a
significant contribution to the distribution function $L$ (see
Appendix 1) In this case the final form of the distribution
function $L$ is given as:

\begin{equation}
\label{2} L\left( {\lambda}  \right) = {\frac{{\sqrt {\pi}} }
{{2\lambda _{0} z}}}{\int\limits_{ - {\frac{{\pi}}
{{2}}}}^{{\frac{{\pi}} {{2}}}} {{\left[ {erf\left(x_{+} \right) -
erf\left( x_{-}\right)} \right]}\cos \theta d\theta}}
\end{equation}

\noindent where: $x_{+}={{\frac{{\lambda - \lambda _{0}}} {{\sigma
\sqrt {2}}} } + {\frac{{\lambda _{0} z}}{{\sigma \sqrt {2}}} }\cos
\theta}$ and $x_{-}={{\frac{{\lambda - \lambda _{0}}} {{\sigma
\sqrt {2}}} } - {\frac{{\lambda _{0} z}}{{\sigma \sqrt {2}}} }\cos
\theta}$

\noindent where:

\noindent $\lambda_{0}$ is the transition wavelength of a spectral
line that arises from a specific point of the equator of a
spherical density region that produces one satellite component, $z
= {\frac{{V_{rot}}} {{c}}}$ ($V_{rot}$ is the rotational velocity
of the specific point) and $erf\left( {x} \right) =
{\frac{{2}}{{\pi}} }{\int\limits_{0}^{x} {e^{ - u^{2}}du}} $ is
the function that describes the Gaussian error distribution.

\section{Discussion and way of application of the proposed model}

Introducing the previous final function of a complex spectral line
(Eq.~\ref{1}), in combination with the distribution function $L$
given in Eq.~\ref{2}, we would like to note the following:

1. The proposed line function (Eq.~\ref{1}) can be used for any
number of absorbing or emitting regions. This means that it can
also be used in the simple case that $i=1$ and $j=0$ or $i=0$ and
$j=1$, meaning when we deal with simple, classical absorption or
emission spectral lines, respectively. This allows us to calculate
all the important physical parameters, such as the rotational, the
radial and the random velocities, the Full Width at Half Maximum,
the Gaussian deviation, the optical depth, the column density
\citep{dan05} and the absorbed or emitted energy, for all simple
and classical spectral lines in all spectral ranges.

2. For each group of the parameters $V_{rot_{i}}$, $V_{rad_{i}}$,
$V_{rand_{i}}$ and $\xi_{i}$, the function $I_{\lambda i} = e^{ -
L_{i} \xi _{i}} $ reproduces the spectral line profile formed by
the $i$ density region of matter, meaning that for each group we
have a totally different profile. This results to the existence of
only one group of $V_{rot_{i}}$, $V_{rad_{i}}$, $V_{rand_{i}}$ and
$\xi_{i}$ giving the best fit of the $i$ component. In order to
accept as the best fit of the observed spectral line,  given by
the groups ($V_{rot_{i}}$, $V_{rad_{i}}$, $V_{rand_{i}}$,
$\xi_{i})$ of all  calculated SACs, one has to adhere to all
physical criteria and techniques, such as:

i) To make a complete identification of  spectral lines in the
region around the studied spectral line and to have the
superposition of the spectral region that we study with the same
region of a classical star of the same spectral type and
luminosity class, in order to identify the existence of spectral
lines that blend with the studied ones, as well as the existence
of SACs.

ii) The resonance lines as well as all  lines originating in a
particular region should have the same number of SACs, depending
on the structure of this region, without influence of ionization
stage or ionization potential of  emitters/absorbers. As a
consequence, the respective SACs should have similar values of
radial and rotational velocities.

iii) The ratio of the optical depths of two resonance lines must be the
same as the ratio of the respective relative intensities.

3. In order to conclude to the group of the
parameters which give  the best fit, we use the model in the
following two steps:

(i) At the first step we consider that the main reason of  line
broadening of the main line and  satellite components is the
rotation of the region which creates the components of the
observed feature and a secondary reason is the thermal Doppler
broadening. This means that we start fitting the line using the
maximum $V_{rot}$. Then we include Doppler broadening, in order to
accomplish the best fit (Rotation case). (ii) At the second step,
we consider the opposite. In this case the main reason of  line
broadening of the main line and  satellite components is supposed
to be the Doppler broadening and the secondary reason is  rotation
of the region which creates  components of the observed feature.
This means that we start fitting the line using the maximal
Doppler broadening. Then we include $V_{rot}$, in order to
accomplish the best fit (Doppler case).

In both  cases (Rotation case and Doppler case) we check the
correct number of satellite components that construct the whole
line profile. At first we fit using the number of components that
give the best difference graph between the fit and the real
spectral line. Then we fit using one component less than in the
previous fit. The F-test between them allows us to take the
correct number of satellite components that construct in the best
way the whole line profile. The F-Test between these two cases
indicates the best way to fit spectral lines. When the F-Test
cannot give definite conclusion on which case we should use, we
still could obtain information about limits of $V_{rot}$ and
$\sigma$. If the F-Test gives similar values, then the rotation
case defines the maximal $V_{rot}$ and the minimal $\sigma$ and
the Doppler case defines the minimal $V_{rot}$ and the maximal
$\sigma$.

4. Profiles of each main spectral line and its SACs are fitted by
the function $e^{ - L_{i} \xi _{i}} $in the case of an absorption
component or $S_{\lambda ej} \left( {1 - e^{ - L_{ej} \xi _{ej}}}
\right)$ in the case of an emission component. These functions
produce symmetrical line profiles. However, most of the spectral
lines  are asymmetric. This fact is interpreted as a systematical
variation of the apparent radial velocities of the density regions
where the main spectral line and its SACs are created. In order to
approximate those asymmetric profiles we have chosen a classical
method. This is the separation of the region, which produces the
asymmetric profiles of the spectral line, into a small number of
sub-regions and each of them is treated as an independent
absorbing shell. In this way we can study the variation of the
density, the radial shift and the apparent rotation as a function
of the depth in every region that produces a spectral line with an
asymmetric profile. All mentioned above have to be taken into
account during the evaluation of our results and one should not
consider that the evaluated parameters of those sub-regions
correspond to independent regions of matter, which form the main
spectral line or its SACs.

5. We suggest that the width of the blue wing is the result of the
composition of profiles of the main spectral line and its SACs.
Thus, the blue wing of each SAC gives the apparent rotational
velocity of the density shell, in which it forms. In order to have
measurements with physical meaning, we should not calculate the
width of the blue wing of the observed spectral feature but the
width of the blue wing of each SAC.

6. We would like to point out that the final criterion to accept
or reject a best fit, is the ability of the calculated values of
the physical parameters to give us a physical description of the
events developing in the regions where the spectral lines
presenting SACs are created.

7. In the proposed distribution function an important factor is $m
= {\frac{{\lambda _{0} z}}{{\sqrt {2} \sigma}} }$. This factor
indicates the kind of the distribution function that is able to
fit in the best way each component's profile.

i) If $m \simeq 3$ we have equivalent contribution of the
rotational and random motions to the line widths.

ii) If $m \simeq 500$ the line broadening is only an effect of the
rotational velocity and the random velocity is negligible. In this
case the profile of the line is the same with the theoretical
profile deriving from the Rotation distribution function.

iii) Finally if $m < 1$ the line broadening is only an effect of
random velocities and the line distribution is the Gaussian.

\section{Testing the model}

\begin{figure}
  \begin{center}
    \FigureFile(75mm,76mm){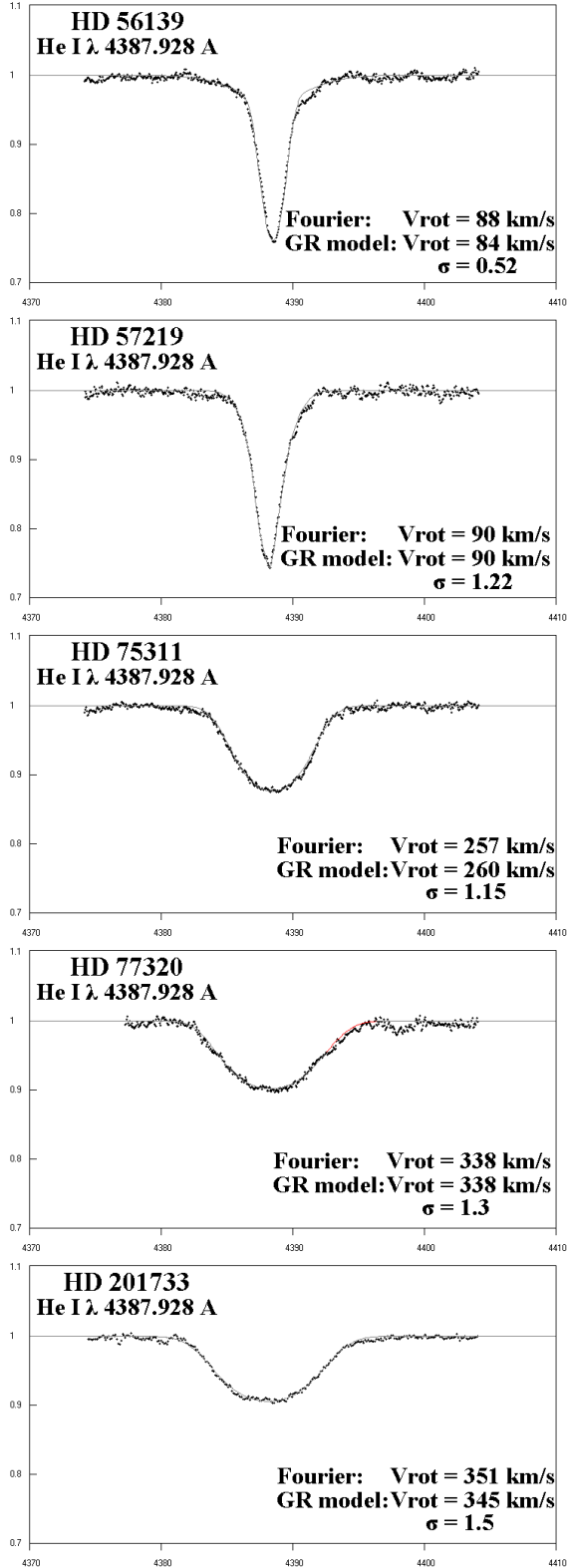}
  \end{center}
  \caption{The five HeI $\lambda$ 4387.928 {\AA} fittings
for the studied Be stars with RG model and the measured rotational
velocities with Fourier analysis and RG model. The differences
between the observed and the reproduced spectral lines are hard to
see, as we have accomplished the best fit.}
\label{HeI}
\end{figure}

In order to check the validity of our model we perform two tests:

I) In order to check the above spectral line function, we
calculated the rotational velocity of He I $\lambda$ 4387.928
{\AA} absorption line for five Be stars, using two methods, the
classical Fourier analysis and our model. In Fig.~\ref{HeI} we
present the five He I $\lambda$ 4387.928 {\AA} fittings for the
studied Be stars and the measured rotational velocities with both
methods. The obtained rotational velocities from our model are in
good agreement with ones obtained with Fourier analysis.

The values of the rotational velocities, calculated with Fourier
analysis, some times, may present small differences compared to
the values calculated with our method, as in Fourier analysis the
whole broadening of the spectral lines is assumed to represent the
rotational velocity. In contrary, our model accepts that a part of
this broadening arises from the random motion of the ions.

We point out that with our model, apart from the rotational
velocities, we can also calculate some other parameters, such as
the standard Gaussian deviation $\sigma$, the velocity of random
motions of emitters/absorbers,  radial velocities of the regions
producing studied spectral lines, the full width at half maximum
(FWHM), the optical depth, the column density and the absorbed or
emitted energy.

II) Additional test of our model is to calculate the random
velocities of  layers that produce the C IV satellite components
of 20 Oe stars with different rotational velocities.

We analyzed the C IV line profiles of 20 Oe stars which spectra
were observed with the IUE - satellite (IUE Archive Search
database\footnote {http://archive.stsci.edu/iue/search.php}). We
examine the complex structure of the C IV resonance lines
($\lambda\lambda$ 1548.155, 1550.774 {\AA}). Our sample includes
the subtypes O4 (one star), O6 (four stars), O7 (five stars), O8
(three stars) and O9 (seven stars). The values of the photospheric
rotational velocities are taken from the catalogue of
\citet{wil63} (see also \citet{ant06}).

In the composite C IV  line profiles we detect two components
in 9 stars, three in 7 stars, four in 3 stars and five in one
star.

\begin{figure}
  \begin{center}
    \FigureFile(30mm,76mm){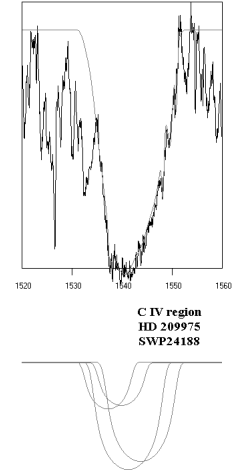}
  \end{center}
  \caption{The C IV resonance lines ($\lambda\lambda$ 1548.155,
1550.774 {\AA}) best fit with GR model for the star HD 209975.
The components obtained from the best fit are shown bottom.}
\label{CIVVrot}
\end{figure}

In Fig.~\ref{CIVVrot} we present the C IV resonance lines best fit
for the star HD 209975.

\begin{figure}
  \begin{center}
    \FigureFile(75mm,76mm){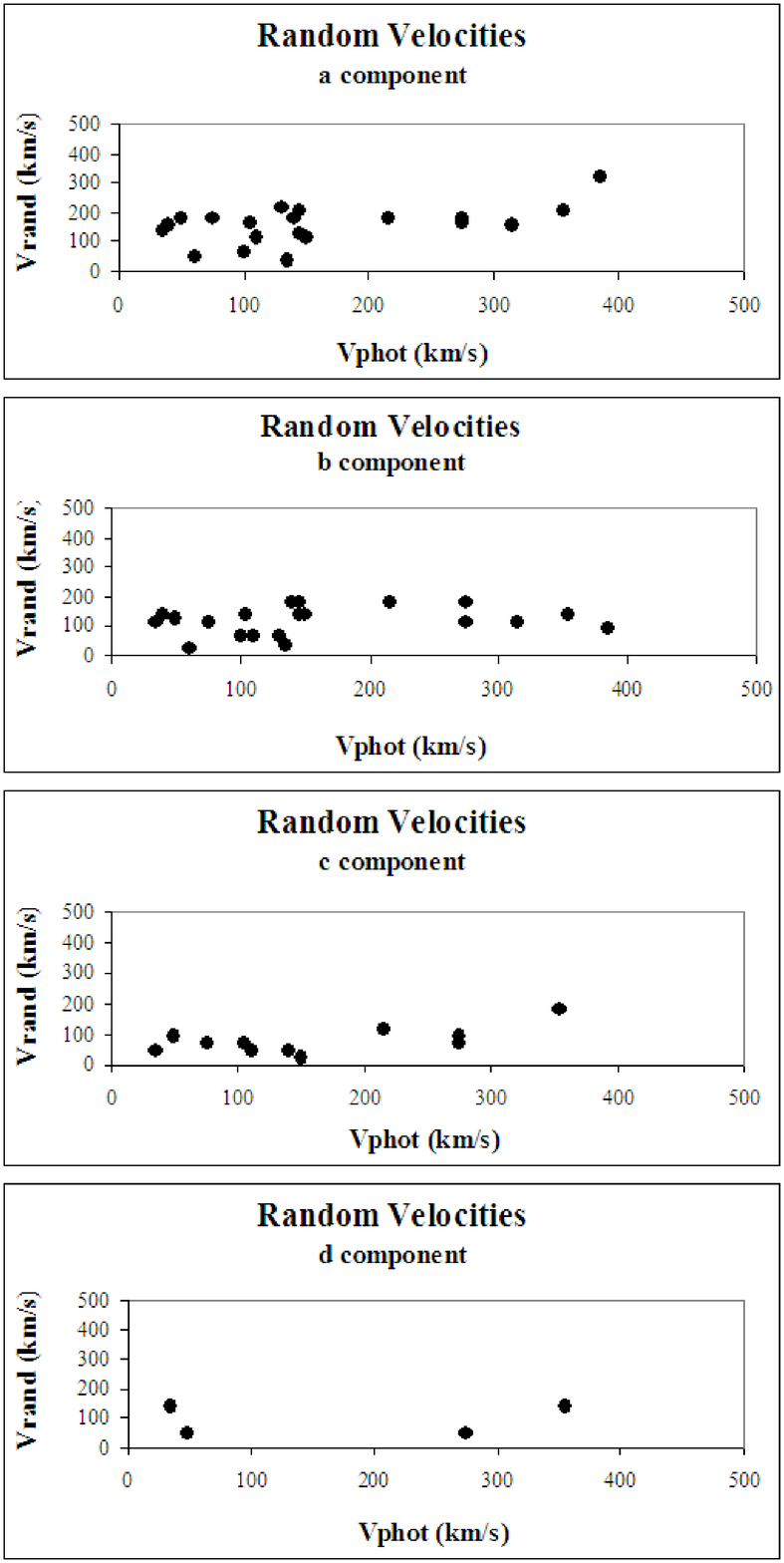}
  \end{center}
  \caption{Random velocities ($V_{rand})$ of the four SACs
as a function of the photospheric rotational velocity ($V_{phot})$
for all the studied Oe stars.} \label{CIVVrand}
\end{figure}

In Fig.~\ref{CIVVrand} we present the random velocities
($V_{rand})$ of each SAC as a function of the photospheric
rotational velocity ($V_{phot})$ for all the studied stars.

As one can see in Fig.~\ref{CIVVrand} the obtained values for
random velocities are in accordance with the classical theory; the
values of the random velocities do not depend on the inclination
angle of the rotational axis. As the ionization potential of the
regions where satellite components are created is the same for all
studied stars, one can expect similar average values of the random
velocities for each absorbing region (here denoted as a-d
component, see Fig.~\ref{CIVVrand}) for all studied stars, as we
obtained  by using the model.

Both of the tests, described above, support our new approach in
investigation of the regions around hot stars, implying that
besides rotation of the regions, one should consider also the
random motion of emitters/absorbers.

\section{Applications of the model: Rotation - density regions vs. photosphere}

Here, we give some examples of application of the model. We  apply
the model to study the complex structure of C IV and Si IV
spectral lines in Oe and Be stars, as well as the radial and
rotational velocities of different components, in order to find
more precise rotational component.

\subsection{C IV density regions  of 20 Oe stars}

\begin{figure}
  \begin{center}
    \FigureFile(75mm,76mm){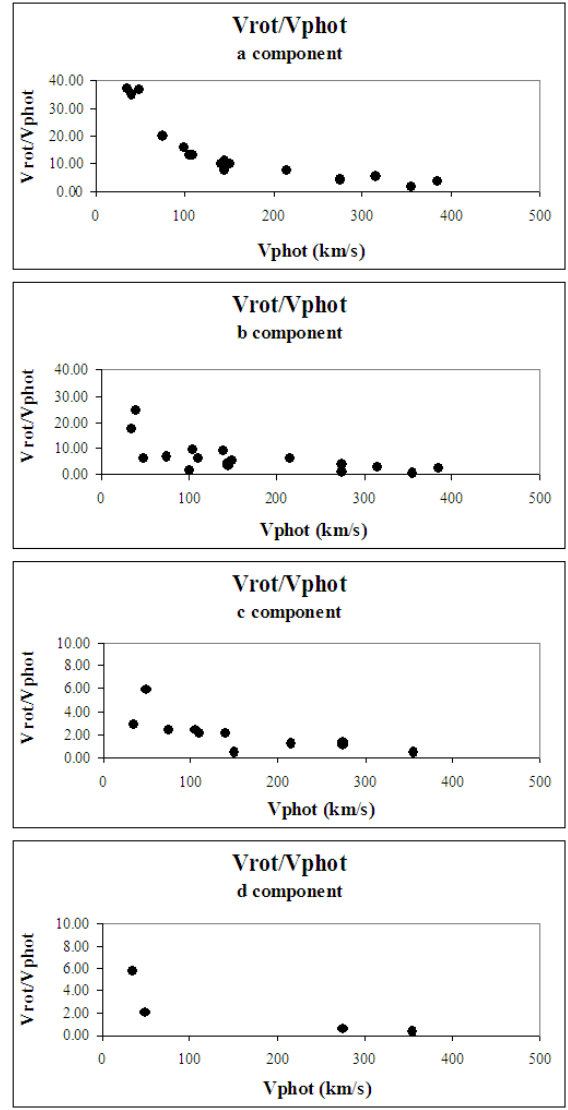}
  \end{center}
  \caption{Ratio $V_{rot}V_{phot}$ of the four SACs as a
function of the photospheric rotational velocity ($V_{phot})$ for
all the studied Oe stars.}
\label{CIVVphot}
\end{figure}

In this application we use the previously mentioned C IV IUE
spectra of the second test. We study the relation between the
ratio $V_{rot}/V_{phot}$ of the first, second, third and fourth
detected component with the photospheric rotational velocity
($V_{phot}$). This ratio indicates how much the rotational
velocity of the specific C IV layer is higher than the apparent
rotational velocity of a star (see also \citet{ant06}). In
Fig.~\ref{CIVVphot} we present our results.

In each region and for each component we can conclude that there
exists an exponential relation between the ratio
$V_{rot}/V_{phot}$ and the photospheric rotational velocity
$V_{phot}$. The maximum ratio $V_{rot}/V_{phot}$ varies from 40,
for the first to 5 for the fourth component (Fig.~\ref{CIVVrot}).
A possible explanation of this situation is the inclination of the
stellar axis.

\subsection{Si IV density regions  of 27 Be stars}

This study is based on the analysis of 27 Be stellar spectra taken
with the IUE - satellite. We examine the complex structure of the
Si IV resonance lines ($\lambda\lambda$ 1393.755, 1402.77 {\AA}).
Our sample includes all the subtypes from B0 to B8. The values of
the photospheric rotational velocities are taken from the
catalogue of \citet{cha01}.

\begin{figure}
  \begin{center}
    \FigureFile(75mm,76mm){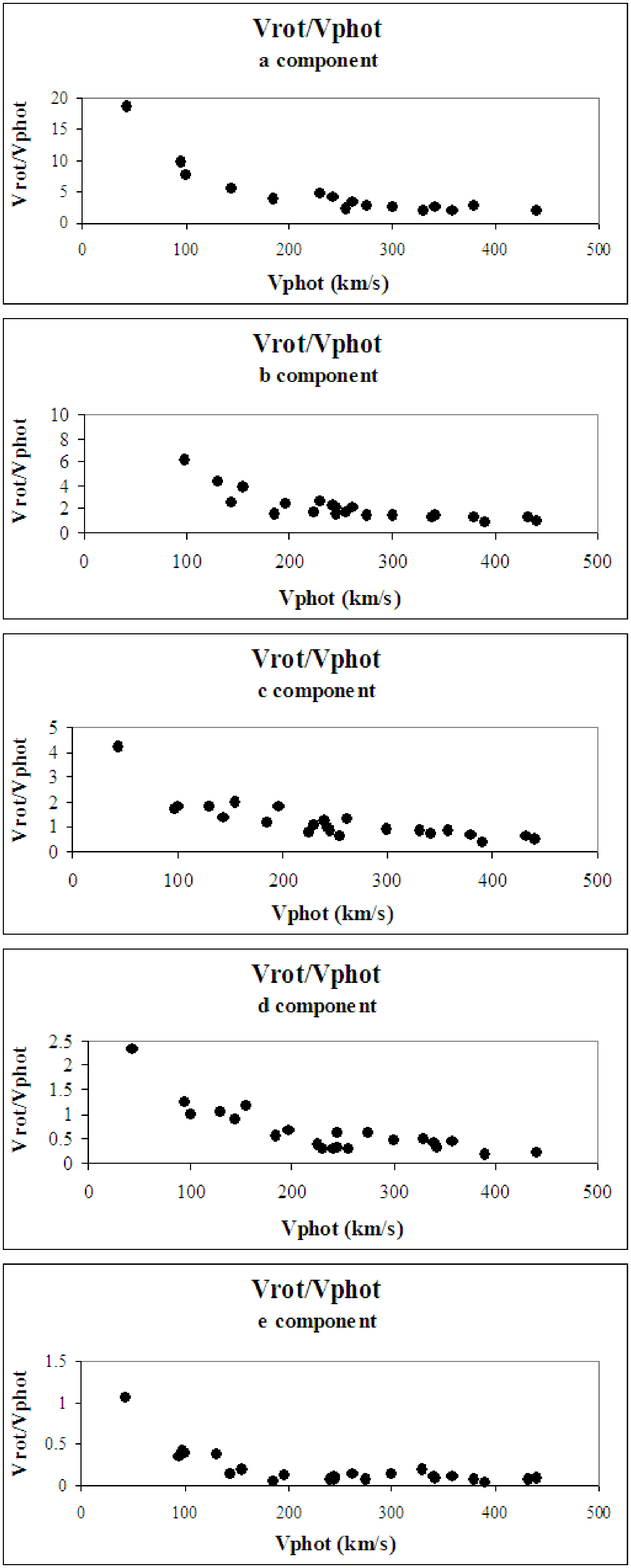}
  \end{center}
  \caption{Ratio $V_{rot}/V_{phot}$ of the five SACs as a
function of the photospheric rotational velocity ($V_{phot})$ for
all the studied Be stars.}
\label{SiIVVrot}
\end{figure}

We found that the Si IV spectral lines consist of three components
in 7 stars, four in 15 stars and five in 5 stars. We study the
relation between the ratio $V_{rot}/V_{phot}$ of the first,
second, third fourth and fifth detected component with the
photospheric rotational velocity ($V_{phot}$). This ratio
indicates how much the rotational velocity of the specific Si IV
layer is higher than the apparent rotational velocity of the star
(see also \citet{lyr06}). In Fig.~\ref{SiIVVrot} we present our
results.

The Si IV resonance lines are composed of three four or five
components. The difference with the case of the C IV resonance
lines in the spectra of 20 Oe stars is that they are composed of
two, three or four components. However, in both cases, in each
region and for each component there exists an exponential relation
between the ratio $V_{rot}/V_{phot}$ and the photospheric
rotational velocity $V_{phot}$. For the satellite components of
the Si IV resonance lines, the maximum ratio $V_{rot}/V_{phot}$
varies from 19, for the first to 1 for the fifth component
(Fig.~\ref{SiIVVrot}). The same phenomenon appears in the case of
the C IV resonance lines in 20 Oe stars, where the maximum ratio
$V_{rot}/V_{phot}$ varies from 40, for the first to 5 for the
fourth component.

\subsection{Rotation in photosphere vs. rotation of density regions}

As one can see in Figs.~\ref{CIVVphot} and~\ref{SiIVVrot}, there
is a good correlation between the rotational velocities of density
regions and photosphere. In both cases (Be and Oe stars) for
rotational velocity of the photosphere $V_{\rm phot}> 200$ km/s,
the rotation of the photosphere correlates with the rotation of
the density regions ($V_{\rm rot}\approx const.\times V_{\rm
phot}$). On the other hand, for $V_{\rm phot}< 200$ km/s, the
ratio $V_{\rm rot}/V_{\rm phot}$ decreases with the rotation in
the photosphere. This can be explained by the inclination effect
and the fact that the density regions are extensive around the
star, i.e. for a high inclination angle the projected photospheric
rotational velocity is small, but in a density region (that lies
extensively around the star), one can detect faster rotation.

\section{Conclusions}

Here we present a new model for fitting the complex UV lines of Be
and Oe stars, where we take into account the possibility that
random motion of ions can significantly contribute to the line
widths. The proposed model can be applied to the spectral
photospherical lines as well as the UV lines originating in the
post-coronal regions. Concerning our work we can give the
following conclusions:

(i) The proposed model can reproduce accurately the complex UV
lines of Be and Oe stars.

(ii) Using the proposed model one can very well separate the
contribution of the rotational and radial velocities in the
density region fitting the complex line profiles.

(iii) The proposed model gives the opportunity to investigate
physical parameters of the regions creating the UV complex line
profiles.

(iv) The proposed model allows us to use the UV photospheric
lines, in order to determine the photospheric rotation.

At the end let us point out that in spite of the fact that today
exists a number of models which are able to reproduce spectra from
stellar atmospheres, there is a problem to find appropriate model
that can fit the complex UV line profiles which are created not
only in the photosphere, but also in the density regions. On the
other hand, the proposed model is able to provide the information
about the physical parameters of density regions as well as
rotation of the photosphere. We hope that the proposed model will
be useful first of all to have first impression about physics of
density layers.

\section{Acknowledgments}

We would like to thank Professor Ryuko Hirata for his very useful
suggestions.

This research project is progressing at the University of Athens,
Department of Astrophysics - Astronomy and Mechanics, under the
financial support of the Special Account for Research Grants,
which we thank very much. The project is co-financed within Op.
Education by the ESF (European Social Fund) and National
Resources. This work also was supported by Ministry of Science and
Environment Protection of Serbia, through the projects: {\it
Influence of collisional processes on astrophysical plasma line
shapes} - P146001 and {\it Astrophysical spectroscopy of
extragalactic objects} - P146002.

\onecolumn
\appendix
\section{Including the random motion in the calculation of L - The Gauss-Rotation model (GR model)}

We consider a spherical shell and a point $A_{i}$ in its equator.
If the laboratory wavelength of a spectral line that arises from
A$_{i}$ is $\lambda _{lab} $, the observed wavelength is: $\lambda
_{0} = \lambda _{lab} \pm \Delta \lambda _{rad}$. If the spherical
density region rotates, we observe a displacement
$\Delta\lambda_{rot}$ and the new wavelength of the center of the
line is $\lambda _{i} = \lambda _{0} \pm \Delta \lambda _{rot}$
where $\Delta \lambda _{rot} = \lambda _{0} z\sin \varphi $, $z =
{\frac{{V_{rot}}} {{c}}}$, $V_{rot} $ is the rotational velocity
of the point A$_{i}$. This means that $\lambda _{i} = \lambda _{0}
\pm \lambda _{0} z\sin \varphi = \lambda _{0} \left( {1\pm z\sin
\varphi}  \right)$ and if $ - {\frac{{\pi }}{{2}}} < \varphi <
{\frac{{\pi}} {{2}}}$ then $\lambda _{i} = \lambda _{0} \left( {1
- z\sin \varphi} \right)$. If we consider that the spectral line
profile is a Gaussian distribution, then $I\left( {\lambda}
\right) = {\frac{{1}}{{\sqrt {2\pi}  \sigma}} }e^{ - {\left[
{{\frac{{\lambda - \kappa}} {{\sigma \sqrt {2}}} }} \right]}^{2}}$
where $\kappa$ is the mean value of the distribution and in the
case of the line profile it indicates the center of the spectral
line that arises from A$_{i}$. This means that:

\[
I\left( {\lambda}  \right) = {\frac{{1}}{{\sigma \sqrt {2\pi}} }
}e^{ - {\left[ {{\frac{{\lambda - \lambda _{0} \left( {1 - z\sin
\varphi} \right)}}{{\sigma \sqrt {2}}} }} \right]}^{2}} =
{\frac{{1}}{{\sigma \sqrt {2\pi}} } }e^{ - {\frac{{{\left[
{\lambda - \lambda _{0} \left( {1 - z\sin \varphi}  \right)}
\right]}^{2}}}{{2\sigma ^{2}}}}}
\]

The distribution function for all the semi-equator is:

\begin{equation}
\label{eq1}I_{1} \left( {\lambda}  \right) = {\int\limits_{ -
{\frac{{\pi }}{{2}}}}^{{\frac{{\pi}} {{2}}}} {{\frac{{1}}{{\sqrt
{2\pi} \sigma}} }e^{ - {\frac{{{\left[ {\lambda - \lambda _{0}
\left( {1 - z\sin \varphi}  \right)} \right]}^{2}}}{{2\sigma
^{2}}}}}\cos \varphi d\varphi}}
\end{equation}

If $\sin \varphi = x$ then $dx = \cos \varphi d\varphi $, $ - 1
\le x \le 1$, and Eq.~\ref{eq1} takes the form

\[
I_{1} (\lambda ) = {\int_{ - 1}^{1} {{\frac{{1}}{{\sigma \sqrt
{2\pi} }}}e^{ -}{{\frac{{[\lambda - \lambda _{0} (1 -
zx)]^{2}}}{{2\sigma ^{2}}}}}dx}}
\]

If we set $u = {\frac{{\lambda - \lambda _{0} \left( {1 - zx}
\right)}}{{\sqrt {2} \sigma}} }$, we have

\[
I_{1} (\lambda ) = {\frac{{1}}{{\lambda _{0} z\sqrt {\pi}
}}}{\int\limits_{{\frac{{\lambda - \lambda _{0} (1 + z)}}{{\sigma
\sqrt {2} }}}}^{{\frac{{\lambda - \lambda _{0} (1 - z)}}{{\sigma
\sqrt {2}}} }} {e^{ - u^{2}}du}}
\]

We consider the function $erf\left( {x} \right) = {\frac{{2}}{{\pi
}}}{\int\limits_{0}^{x} {e^{ - u^{2}}du}} $. It is the known
function which describes the Gaussian error distribution.

If we take into account this function, $I_{1} (\lambda )$ takes
the form

\[
{\rm I}_{1} (\lambda ) = {\frac{{1}}{{\lambda _{0} z\sqrt {\pi}} }
}{\left[ {{\int\limits_{0}^{{\frac{{\lambda - \lambda _{0} (1 -
z)}}{{\sigma \sqrt {2}}} }} {e^{ - u^{2}}du -
{\int\limits_{0}^{{\frac{{\lambda - \lambda _{0} (1 + z)}}{{\sigma
\sqrt {2}}} }} {e^{ - u^{2}}du}}} }}  \right]}
\]

\[
{\rm I}_{1} \left( {\lambda}  \right) = {\frac{{1}}{{\lambda _{0}
z\sqrt {\pi}} } }{\left[ {{\frac{{\pi}} {{2}}}erf\left(
{{\frac{{\lambda - \lambda _{0} (1 - z)}}{{\sigma \sqrt {2}}} }}
\right) - {\frac{{\pi }}{{2}}}erf\left( {{\frac{{\lambda - \lambda
_{0} (1 + z)}}{{\sigma \sqrt {2}}} }} \right)} \right]}
\]

Thus, we finally have

\[
{\rm I}_{1} \left( {\lambda}  \right) = {\frac{{\sqrt {\pi}} }
{{2\lambda _{0} z}}}{\left[ {erf\left( {{\frac{{\lambda - \lambda
_{0} \left( {1 - z} \right)}}{{\sigma \sqrt {2}}} }} \right) -
erf\left( {{\frac{{\lambda - \lambda _{0} \left( {1 + z}
\right)}}{{\sigma \sqrt {2}}} }} \right)} \right]}
\]

\noindent and the distribution function from the semi-spherical
region is:

\begin{equation}
\label{eq2} {\rm I}_{final} \left( {\lambda}  \right) =
{\frac{{\sqrt {\pi} }}{{2\lambda _{0} z}}}{\int\limits_{ -
{\frac{{\pi}} {{2}}}}^{{\frac{{\pi }}{{2}}}} {{\left[ {erf\left(
{{\frac{{\lambda - \lambda _{0}}} {{\sigma \sqrt {2}}} } +
{\frac{{\lambda _{0} z}}{{\sigma \sqrt {2}}} }\cos \theta} \right)
- erf\left( {{\frac{{\lambda - \lambda _{0}}} {{\sigma \sqrt {2}}}
} - {\frac{{\lambda _{0} z}}{{\sigma \sqrt {2}}} }\cos \theta}
\right)} \right]}\cos \theta d\theta}}
\end{equation}

\noindent (Method Simpson)

In Eq.~\ref{eq2}, from $\lambda_{0}$ we calculate the value of the
radial velocity ($V_{rad}$), from $z$ we calculate the rotational
velocity ($V_{rot}$) and from $\sigma$ we calculate the random
velocity ($V_{rand}$).

This distribution function ${\rm I}_{final} \left( {\lambda}
\right)$ has the same form with the distribution function of the
absorption coefficient $L$ and may replace it in the line
functions $e^{ - L\xi} $ or $S_{\lambda e} \left( {1 - e^{ -
L_{ej} \xi _{ej}}}  \right)$, in the case when the line broadening
is an effect of both the rotational velocity of the density region
as well as the random velocities of the ions. This means that now
we have a new distribution function to fit each satellite
component of a complex line profile that presents DACs or SACs. We
name this function Gauss-Rotation distribution function (GR
distribution function).

\twocolumn

\end{document}